\documentstyle[12pt]{article}

\renewcommand{\baselinestretch}{1.176}

\catcode`\@=11
\def\capfont@figure{\rm\small}
\def\fnum@figure{{\capfont@figure Fig. \thefigure}}

\renewcommand\section{\vspace{2\baselineskip}\@startsection{section}
{1}{\z@}{-3pt}{+3pt}{\centering\normalsize}}
\catcode`\@=12

\newcommand{\dib}[2]{\begin{figure}\vskip#1\caption[]{\small #2}
\end{figure}}

\newcommand{\rmi}{\mbox{\rm I}}
\newcommand{\om}{\omega}
\newcommand{\gb}{\beta}
\newcommand{\ep}{\epsilon}

\newcommand{\fr}{\frac{1}{2}}
\newcommand{\pat}{\partial}

\newcommand{\diag}{{\rm diag}}
\newcommand{\const}{{\rm const}}

\begin{document}

\vspace*{2.5cm}
\begin{center}
\LARGE\bf Chaos and Quantum Chaos in\\ Cosmological Models
\end{center}

\vspace{0.7cm}
\centerline{\bf R.~GRAHAM}

\vspace{1.4cm}
\centerline{\small Fachbereich Physik, Universitaet GH Essen, Germany}

\vspace{1.8cm}

\parbox{16cm}
{{\bf Abstract} - Spatially homogeneous cosmological models reduce to
Hamiltonian systems in a low dimensional Minkowskian space moving on
the total energy shell $H=0$. Close to the initial singularity some
models (those of Bianchi type VIII and IX) can be reduced further, in
a certain approximation, to a non-compact triangular billiard on a
2-dimensional space of constant negative curvature with a separately
conserved positive kinetic energy. This type of billiard has long
been known as a prototype chaotic dynamical system. These facts are
reviewed here together with some recent results on the energy level
statistics of the quantized billiard and with direct explicit
semi-classical solutions of the Hamiltonian cosmological model to
which the billiard is an approximation. In the case of Bianchi type
IX models the latter solutions correspond to the special boundary
conditions of  a `no-boundary state' as proposed by Hartle and
Hawking and of a `wormhole' state.}

\section*{\bf INTRODUCTION}
The subject of chaos in cosmology is really an old one and predates
the intense modern interest in chaotic dynamical systems. It started
with the classical work of Belinsky et al~1969 \cite{A} and of
Misner~1969 \cite{B} who found that the evolution of certain
homogeneous cosmological models back into the past towards the `Big
Bang' shows an oscillatory behavior, some aspects of which in modern
terminology one has to call chaotic. Useful references for these
cosmological models are the books by Ryan~1972 \cite{C} and Ryan and
Shepley~1975 \cite{D}. The chaotic aspects of their evolution have
been examined in numerous papers starting with the review by
Barrow~1982 \cite{E}. Khalatnikov et al~1985 \cite{F} give a detailed
discussion. A dynamical systems approach was pioneered by
Bogoyavlensky and Novikov~1973 \cite{G} and elaborated by Ma and
Wainwright~1989 \cite{H}. Among the more recent works on this subject
those of Pullin~1990 \cite{I} and Rugh~1992 \cite{J} may be
mentioned. Most recently a Painlev\'e analysis was applied to this
dynamical system by Cantopoulos et al~1993 \cite{K} who found,
surprisingly, that the system satisfies the Painlev\'e conditions
which are necessary (but not sufficient) for a system to be
integrable.

Early on, again before the explosion of the
interest in chaos took place, Misner~1972 \cite{L} in a seminal paper
and his student Chitre~1972 \cite{M} in his thesis introduced
coordinates in which the most interesting homogeneous cosmologies,
those of Bianchi types VIII and IX, sufficiently close to the
singularity reduce to a special two-dimensional billiard on a
homogeneous space of constant negative curvature.  The study of this
kind of billiards is an even more ancient subject in mathematics,
dating back to the work of Hadamard~1898 \cite{N}, Artin~1924
\cite{O} and Hopf~1937 \cite{P}. An extensive review is due to Balazs
and Voros~1986 \cite{Q}.

Given a dynamical system of Hamiltonian form the temptation to
quantize seems to be irresistable. Hence the field of quantum chaos,
which is concerned with the quantization of classically chaotic
systems with the aim of understanding features of chaos and quantum
mechanics, hence also the field of quantum cosmology, which starts
with the quantization of spatially homogeneous cosmological models
with the aim to achieve an understanding of quantum features
of cosmology. For the
Bianchi type IX cosmology this was first done by Misner~1972 in
\cite{L}.  However, the present interest in quantum cosmology was
mainly triggered by the proposal of a specific initial condition,

the `no-boundary-state', by Hawking~1982 \cite{R}, and Hartle and
Hawking~1983 \cite{S}. Quantized versions of the homogeneous Bianchi
type IX cosmological models were also investigated in this context,
in semi-classical WKB-type approximations, e.g. in papers by Hawking
and Luttrell~1984 \cite{T}, and Moss and Wright~1985 \cite{U} who
study consequences of the `no-boundary'condition and by Del Campo and
Vilenkin~1989 \cite{V}, and Graham and Sz\'epfalusy~1990 \cite{W} who
instead pose the boundary condition of an outgoing wave. From the
point of view of quantum chaos, the Bianchi IX model has also been
studied further in papers by Furusawa~1986 \cite{X} and Berger~1989
\cite{Y} who study the dynamics of wave-packets and by Graham et
al.~1991 \cite{Z} and Csord\'as et al.~1991 \cite{a} who study level
statistics in Misner's billiard limit of the model. More recently
further detailed studies of a class of billiards including the
special one relevant to the Bianchi VIII and IX model have appeared
in the quantum chaos literature (Bogomolny et al.~1992 \cite{b,c},
Bolte et al.~1992 \cite{d}, Schmit~1991 \cite{e} Eisele and
Mayer~1993 \cite{f}).

Finally, as the most recent development, exact solutions of quantized
Bianchi IX models were found independently by Moncrief and Ryan~1991
\cite{g}, Graham~1991 \cite{h}, and Bene and Graham~1993 \cite{i}. It
is now understood that the solution found in the first two references
describe virtual quantum wormholes (D.Eath~1993 \cite{j}, Bene and
Graham~1993 \cite{i}), while a further exact solution by Bene and
Graham ~1993 \cite{i} gives the `no-boundary-state' of the model in
a simple analytical form.

The present paper has the purpose to review some of these results. In
the next section we establish the necessary background on spatially
homogeneous cosmological models. Then we proceed to consider Misner's
approximate asymptotic reduction of the Bianchi VIII and Bianchi IX
models to 2-dimensional billiards on a space of constant negative
curvature, and discuss some results following from this reduction.
Next, the Bianchi VIII and IX models are quantized, giving rise to
their Schr\"odinger equation, called Wheeler DeWitt equation in the
present context. We then discuss some of the properties of the
billiard approximation of the quantized model. Finally, we turn to
exact solutions of the Wheeler DeWitt equation and their physical
interpretation.

\section*{\bf SPATIALLY HOMOGENEOUS COSMOLOGICAL MODELS}

We consider the class of space-times such that through every point
there passes a space-like 3-manifold which is left invariant by the
actions of a three-dimensional Lie group (see e.g. Ryan and
Shepley~1975 \cite{D}). Such space-times are called spatially
homogeneous. The three-dimensional Lie groups have been classified by
Bianchi into nine different types (`Bianchi types'). They can be
distinguished by the Lie algebra of the group-generators, called
`Killing vectors' in general relativity. It is convenient to use an
invariant (non-coordinate) basis and its dual of basis 1-forms
$\omega^i$ on the invariant 3-manifold.

The space-time metric then takes the general form
\begin{equation}
ds^2=-dt^2+g_{ij}(t)\om^i\om^j
\label{eq:2-1}
\end{equation}
where $dt$ is the time-like one-form orthogonal to the invariant
3-manifold and $t$ is the standard cosmic time coordinate. The
components $g_{ij}(t)$ of the 3-metric then depend only on $t$. The
basis 1-forms satisfy
\begin{equation}
d\om^i=\fr {C^i}_{jk}\om^j\wedge\om^k
\label{eq:2-2}
\end{equation}
(we recall the usual definition
$\om^j\wedge\om^k=\om^j\bigotimes\om^k-\om^k\bigotimes\om^j$) where
the ${C^i }_{jk}$ are the structure constants of the homogeneity
group. Of special interest here will be the Bianchi type VIII with
\begin{equation}
-{C^1}_{23}=+{C^1}_{32}=+{C^2}_{31}=-{C^2}_{13}=+
   {C^3}_{12}=-{C^3}_{21}=1
\label{eq:2-3}
\end{equation}
and in particular the Bianchi type IX with
\begin{equation}
{C^i}_{jk}=\ep_{ijk}.
\label{eq:2-4}
\end{equation}
In the latter case the homogeneity group is $SO(3)$, the invariant
3-manifold is compact and topologically a 3-sphere. This case
therefore includes the closed Friedmann-Robertson-Walker (FRW)
space-time and all homogeneous anisotropic modifications thereof. The
invariant 3-manifolds of Bianchi-type VIII are non-compact and
describe open space-times. (The open FRW space-times are not of this
particular Bianchi-type, however, see Ryan and Shepley~1975
\cite{D}).

We shall restrict the further discussion to diagonal models, where
$g_{ij}(t)$ can consistently be chosen diagonal,
\begin{equation}
g_{ij}=\frac{1}{6\pi}\diag \left(
 e^{(2\alpha+2\beta_++2\sqrt{3}\beta_-)},
   e^{(2\alpha+2\beta+-2\sqrt{3}\beta_-)},
    e^{(2\alpha-4\beta_+)}\right)
\label{eq:2-5}
\end{equation}
with three time-dependent parameters $\alpha$, $\beta_+$, $\beta_-$.
The special space-time metric (\ref{eq:2-1}) can now be inserted in
the action functional of general relativity, in particular in its
first-order (Hamiltonian) form first given by Arnowitt, Deser, and
Misner~1962 \cite{k}. We shall neglect the matter contribution in the
action, which, in the anisotropic case,  can be shown to become
unimportant close to the initial singularity. The result is a reduced
action of the form
\begin{equation}
S=\int\,d\lambda\left(p_\nu\dot{q}^\nu-N(\lambda)H(q,p)\right)
\label{eq:2-6}
\end{equation}
with
\begin{eqnarray}
  N(\lambda)d\lambda &=& \sqrt{\frac{3\pi}{2}}e^{-3\alpha(t)}dt
  \nonumber\\
  \dot{q}^\nu &=&\frac{dq^\nu}{d\lambda}\\
     H(q,p) &=& \fr(-p_\alpha^2+p_+^2+p_-^2)+
                  V(\alpha,\gb_+,\gb_-).\nonumber
   \label{eq:2-7}
\end{eqnarray}
Here $N(\lambda)>0$ is a Lagrange parameter called the lapse
function, whose variation enforces the condition
\begin{equation}
H=0
\label{eq:2-8}
\end{equation}
$q^\nu$, $p_\nu$ are used to denote the coordinates in the space of
diagonal, homogeneous 3-metrics (mini-superspace)
$\alpha,\gb_+,\gb_-$ and their canonical momenta. The action
(\ref{eq:2-6}) has the same form as for a particle in a 3-dimensional
potential well, except for the Minkowskian nature of the metric in
mini-superspace. The potential $V(\alpha,\gb_+,\gb_-)$ is given by
the scalar curvature $^{(3)}R$ of the homogeneous 3-metric
\begin{equation}
V=-12\pi^2\det (g_{ij})^{(3)}R.
\label{eq:2-9}
\end{equation}
For the two Bianchi-types of special interest here it is given by
(Ryan and Shepley~1975 \cite{D})
\begin{equation}
  \begin{array}{l}
V=\frac{1}{6} e^{4\alpha}\Bigg[
    2e^{4\gb_+}(\cosh (4\sqrt{3}\gb_-)-1)+e^{-8\gb_+}\\
\hspace{2.4cm} \pm 4e^{-2\gb_+}\cosh(2\sqrt{3}\gb_-)\Bigg]
   \end{array}
\label{eq:2-10}
\end{equation}
where the upper and lower sign apply to Bianchi-type VIII and
Bianchi-type IX, respectively.

The action (\ref{eq:2-6}) with eqs. (\ref{eq:2-7})-(\ref{eq:2-10})
correctly generates the reduced equations of motion obtained from
Einstein's equations with the ansatz (\ref{eq:2-5}). It therefore
gives an exact description of the spatially homogeneous models.

We note that the potential (\ref{eq:2-10}) for Bianchi-type VIII is
positive everywhere and decreases towards $0$ for $\gb_-=0$,
$\gb_+\rightarrow\infty$. It is therefore like a space-time dependent
mass of a relativistic particle with $\alpha$ playing the role of the
time-coordinate changing monotoneously with $\lambda$. For
Bianchi-type IX, on the other hand, the potential is everywhere
increasing for $|\gb_+|\rightarrow\infty$ or
$|\gb_-|\rightarrow\infty$, and $V$ becomes {\bf negative} in a
neighborhood of the origin $\gb_+=\gb_-=0$.  Therefore the action of
$V$ cannot be compared to a mass-term in this case, and $\alpha$ need
not change monotoneously with $\lambda$.

However, it can be shown (Bogoyavlensky and Novikov~1973 \cite{G},
Rugh~1992 \cite{J}) that $\alpha$ cannot have a minimum
$\frac{d\alpha}{d\lambda}=0$, $\frac{d^2\alpha}{d\lambda^2}>0$ and
can have only a single maximum. To see this we use $H=0$ (and $N=1$)
to express
\begin{equation}
  -\frac{d^2\alpha}{d\lambda^2}=-\frac{\pat H}{\pat\alpha}=
    -4V=-2\left(\frac{d\alpha}{d\lambda}\right)^2+
    2p_+^2+2p_-^2
\label{eq:2-11}
\end{equation}
which implies $d^2\alpha/d\lambda^2<0$ (except for $p_+=p_-=0$)for
$d\alpha/d\lambda=0$.  Hence, sufficiently close to the singularity
$\alpha\rightarrow-\infty$, where a maximum of $\alpha$ does not
occur, $\alpha$ or any monotoneous function of $\alpha$ can in fact
be used as a time-coordinate.

For $\alpha\rightarrow -\infty$ the potential (\ref{eq:2-10})
approaches zero for values of $\gb_+,\gb_-$ inside a triangular
region of the $(\gb_+,\gb_-)$-plane bounded by the three straight
lines
\begin{equation}
  \begin{array}{l}
  \alpha+\gb_+\pm\sqrt{3}\gb_-=0\\
  \alpha-2\gb_+=0. \end{array}
\label{eq:2-12}
\end{equation}
The three sides of this triangle recede with decreasing `time'
$\alpha$ with velocity $1/2$. The velocity of the system point with
respect to the time $\alpha$ in the region where $V$ is negligible is
approximately 1. The potential (\ref{eq:2-10}) rises steeply
(exponentially) outside the lines (\ref{eq:2-12}) with the exception
of narrow channels reaching to infinity. These channels are located
symmetrically at all three corners of the triangle for Bianchi-type
IX and in them the potential approaches zero from below for
$\alpha\rightarrow -\infty$. For Bianchi-type VIII there is only one
channel at the corner $\gb_-=0$, $\gb_+=-\alpha$ reaching to
$\gb_+=+\infty$, and there the potential approaches zero from above.
In an apparently reasonable approximation in the limit
$\alpha\rightarrow -\infty$ the exponentially rising potential walls
can be replaced by the infinitely steep walls of a billiard located
at the straight lines (\ref{eq:2-12}). The picture of a billiard ball
of unit velocity elastically bouncing from the walls of an
equilateral triangle receding with velocity $1/2$ permits a thorough
and lucid qualitative discussion of the resulting dynamics, which has
been given in many papers and textbooks, starting with the classic
works of Belinski et al~1969 \cite{A} and Misner~1969 \cite{B}.
Detailed numerical simulations fully confirming the qualitative
picture have recently been reported e.g. by Rugh~1992 \cite{J}. We
refer to the quoted literature for details. The channels at the
corners, and the (for $\alpha\rightarrow-\infty$ increasingly)
unlikely events, where the system point enters one of those channels,
are not described by the billiard approximation. This fact makes it
hard to assess the quality of the billiard approximation. In
particular, chaos in the billiard approximation need not imply chaos
in the unapproximated system, whose presence or absence is, in fact,
still an open question (see Cantopoulos et al~1993 \cite{K} for an
indication that the system might be actually integrable; however, it
should be noted that eq. (\ref{eq:2-11}) e.g. implies that the
phase-space function $p_{\alpha}e^{-2\alpha}$ varies monotonously
with $\lambda$ which makes the system very special indeed, and calls
for extreme caution before drawing too strong conclusions from the
positive result of the Painlev\'e test for integrability).

In the next two sections we shall be concerned with the properties of
the cosmological models in the billiard-approximation. However, in
the final section we shall return to the description by the full
Hamiltonian (8). The `free'motion between two bounces is
called a `Kasner epoch' because the Kasner universe is described by
the Hamiltonian (8) with $V\equiv 0$ (Ryan and
Shepley~1975 \cite{D}). Repeated bounces between the same two walls
constitute an `era'. For $\alpha\rightarrow -\infty$ , the motion
then consists of an infinite sequence of `eras' following each other
in which the two walls defining each era are permutated in a chaotic
manner (Barrow~1982 \cite{E}, Khalatnikov et al.~1985 \cite{F}).

\section*{\bf REDUCTION TO A BILLIARD ON THE PSEUDO-SPHERE}

Discussion of chaos in the triangular billiard with moving walls are
hampered by the fact that, with the use of $\alpha$ as an effective
time-parameter, the system is explicitely time-dependent
and the long-time limit $\alpha\rightarrow -\infty$, depending on its
precise definition, is either trivial (free particle) or it does not
exist (the particle-bounces never disappear). In fact an obvious but
important point emphasized by Rugh~1992 \cite{J} and Pullin~1990
\cite{I} is the proper choice of the time-coordinate in discussions
of chaos in general relativity: as Lyapunov exponents and
Kolmogorov-Sinai entropy are not invariant under the free
redefinitions of the time-coordinate permitted by general relativity
some suitable coordinate condition must be posed in addition in order
to preserve the usefulness of these mathematical concepts. Obviously
one useful choice is a time variable which eliminates the explicit
time-dependence at least asymptotically for sufficiently long time.

For the problem at hand such coordinates were provided by
Misner~1972 \cite{L}, and first put to use in Chitre's thesis~1972
\cite{M}. These coordinates $(t,\xi,\phi)$ are defined by
\begin{eqnarray}
\gb_+ &=& e^t\sinh \xi\cos\phi\nonumber\\
\gb_- &=& e^t\sinh\xi\sin\phi\\
\alpha &=& -e^t\cosh\xi\quad (\alpha<0).\nonumber
\label{eq:3-1}
\end{eqnarray}
Here $t$ is not to be confused with the synchronous time-coordinate
in eq.(\ref{eq:2-1}). Rather, in the Minkowskian metric of the
Hamiltonian (8) it is a time-like coordinate like $\alpha$. In fig.~1
we give a schematic plot illustrating the coordinate change. The
triangular billiard walls (\ref{eq:2-12}) are shown in the
$(\gb_+,\gb_-)$-plane for fixed $\alpha<0$.  The corners of the
triangle are on the circle $\gb_+^2+\gb_-^2=\alpha^2$.  A fixed value
of $\xi$ corresponds to the smaller circle
$\gb_+^2+\gb_-^2=\alpha^2\tanh^2\xi$. The corners of the triangle are
therefore at $|\xi|=\infty$ in the new coordinates. The narrow
potential channels at the corner of the Bianchi VIII and at all three
corners of the Bianchi IX potentials can therefore not even be
described in the new coordinates.

The billiard walls (\ref{eq:2-12}) in the new coordinates are at
\begin{equation}
 \tanh\xi = -\fr \sec\left(\phi+m\frac{2\pi}{3}\right)
\label{eq:3-2}
\end{equation}
\dib{8cm}{Billiard walls (12) and a circle $\xi=\const$ in the
$(\gb_+,\gb_-)$-plane for fixed $\alpha$.}
with $m=0,\pm 1$. They are independent of the new variable $t$ which
will be used as a new time coordinate instead of $\alpha$. The
Hamiltonian (\ref{eq:2-7}) expressed in the new coordinates and their
canonical momenta takes the form
\begin{equation}
 H=\fr\left(-p_t^2+p_\xi^2+\frac{p_\phi^2}{\sinh^2\xi}\right)e^{-2t}+
  V(\xi,\phi,t).
\label{eq:3-3})
\end{equation}
Replacing the lapse function $N(\lambda)$ by
\begin{equation}
N(\lambda)\rightarrow N(\lambda)e^{2t}
\label{eq:3-4}
\end{equation}
the Hamiltonian becomes
\begin{equation}
 H=\fr\left(-p_t^2+p_\xi^2+\frac{p_\phi^2}{\sinh^2\xi}\right)
  +e^{2t}V(\xi,\phi,t).
\label{eq:3-5}
\end{equation}
It still has to satisfy the condition $H=0$. For $t\rightarrow\infty$
the potential approaches 0 like $\exp(-{\mbox{\rm const.}}e^t+2t)$,
const. $>0$, inside the billiard walls (\ref{eq:3-2}), and it
approaches infinity outside these walls. Therefore the Hamiltonian
(\ref{eq:3-5}) ceases to be time-dependent in that limit and
the canonical momentum $p_t$ is asymptotically conserved. The motion
in this limit therefore becomes that of a free particle moving with
arbitrary conserved energy $\fr p_t^2$ on a 2-dimensional space with
metric
\begin{equation}
  dl^2=d\xi^2+\sinh^2\xi d\phi^2
\label{eq:3-6}
\end{equation}
and bouncing elastically from the infinitely steep walls at the lines
(\ref{eq:3-2}). The Riemannian space defined by eq. (\ref{eq:3-6})
has constant negative curvature, with Gaussian curvature $K=-1$ and
Riemannian scalar curvature $R=-2$. Replacing $\xi$ by $r=\tanh\xi/2$
it is represented by the pseudo-sphere $r\le 1$ with
\begin{equation}
  dl^2=\frac{dr^2+r^2d\phi^2}{1-r^2}
\label{eq:3-7}
\end{equation}
or by the Poincar\'e half-plane $Im z\ge 0$ with
\begin{eqnarray}
z &=& \frac{\sqrt{3}}{2}i
   \frac{1-re^{i(\phi+\frac{\pi}{6})}}{1+re^{i(\phi+\frac{\pi}{6})}}
     -\fr=x+iy\nonumber\\
  dl^2 &=& \frac{dx^2+dy^2}{y^2}.
\label{eq:3-8}
\end{eqnarray}
It is obvious from eqs. (\ref{eq:3-7}), (\ref{eq:3-8}) that the
metric in these coordinates is conformal to a Euclidean metric, i.e.
all angles are as in a Euclidean space. The walls (\ref{eq:3-2}) are
geodesics of the metric and define a triangular billiard with corners
at infinity.  Therefore the billiard is non-compact. However, its
total metric area is finite. In the coordinates (\ref{eq:3-8})
the walls are at
\begin{equation}
  x=-1,\, x=0,\, y\ge 0;\, \left(x+\fr\right)^2+y^2=\frac{1}{4},\,
y\ge 0.
\label{eq:3-9}
\end{equation}
In these coordinates geodesics are semi-circles in the upper half of
the $z$-plane with centers on the real
axis. Due to the metric points on the real axis and points with
$y=\infty$ are at infinity.
For many purposes a geodesic which is repeatedly reflected on the
walls is more conveniently represented on a somewhat enlarged domain,
again bounded by geodesic walls, which are identified in opposite
pairs corresponding to periodic boundary conditions. The smallest
such domain preserving the geodesic is called the fundamental domain.
It is shown in fig.~2 where the original billiard is displayed
together with its extension to a fundamental domain. The letters a
and b indicate which sides are to be identified. The conformal
transformations identifying opposite sides together with the unit
transformation generate elements of the `Fuchsian group' of the
billiard. Explicitely, these transformations are
\begin{equation}
  A:\, z'=z+2;\quad B:\, z'=\frac{z}{2z+1}
\label{eq:3-10}
\end{equation}
and their inverse. In order to return from the fundamental domain of
the Fuchsian group to the original domain of the billiard, points
connected by a reflection on the wall $x=0 (z'=-z^*)$ are identified.

Chitre~1972 \cite{M} first used this construction to demonstrate
ergodicity and mixing of the cosmological billiard invoking theorems
by Hopf~1936 \cite{P} and Hedlund~1939 \cite{k} on geodesic flows on
the quotient space of the pseudo-sphere and a Fuchsian group. For a
similar non-compact billiard whose fundamental domain is the modular
domain $|z|>1$, $-\fr<x<\fr$, $y>0$ with $A:z'=z+1$, $B:z'=-1/z$
Artin~1924 \cite{O} already gave a demonstration of ergodicity. He
used the theory of continued fractions to show that the
transformations of the Fuchsian group, applied to the real footpoints
of a geodesic semi-circle, generate a sequence of footpoints which
lie dense on the real axis (and hence generate semi-circles which lie
dense in the modular domain) except for a set of footpoints of
measure $0$. The transformations (\ref{eq:3-10}) restricted to the
real axis play an equally important role in the quantized billiard
(see below).
\dib{9cm}{The cosmological triangular billiard (full
lines) in the complex $z$-plane and its extension to a fundamental
domain (dashed lines) with sides $a, a'=A(a)$ and $b,b'=B(b)$
pairwise identified. }

Due to the fact that the walls are geodesics the Lyapunov exponent
$\lambda$ and the Kolmogorov-Sinai entropy $h$
immediately follow from the equations for the geodesic deviations and
are given by $\lambda=h=|p_t|$.

\section*{\bf QUANTIZATION}

The canonical quantization of the Hamiltonian (7) with the
condition (\ref{eq:2-7}) yields the Wheeler DeWitt equation
\begin{equation}
 \Bigg\{\fr\bigg((\frac{\pat}{\pat\alpha})^2-(\frac{\pat}
                 {\pat\gb_+})^2-
                 (\frac{\pat}{\pat\gb_-})^2\bigg)+
         V(\alpha,\gb_+,\gb_-)\Bigg\} \psi(\alpha,\gb_+,\gb_-)=0.
\label{eq:4-1}
\end{equation}
Eq. (\ref{eq:4-1}) is particularly simple because the lapse function
was chosen so as to make the metric in the variables
$(\alpha,\gb_+,\gb_-)$ flat.  For a different choice of the lapse
function the metric is no longer flat and its scalar curvature should
appear in (\ref{eq:4-1}) together with the Laplace-Beltrami operator
to render (\ref{eq:4-1}) invariant under such conformal
transformations (Misner~1972 \cite{L}). In the present section we
wish to consider eq. (\ref{eq:4-1}) in the billiard approximation
where it reduces to the quantized version of eq. (\ref{eq:3-3})
(whose lapse function is still the same as in (\ref{eq:4-1})) with
infinitely steep potential walls, i.e.
\begin{equation}
  e^{-2t}\Bigg[
     e^{-t}\frac{\pat}{\pat t}\left(e^{-t}\frac{\pat}{\pat t}\right)
    -\frac{1}{\sinh\xi}\frac{\pat}{\pat\xi}
      \left(\sinh\xi\frac{\pat}{\pat\xi}\right)
    -\frac{1}{\sinh^2\xi}\frac{\pat^2}{\pat\phi^2}\Bigg]\psi=0
\label{eq:4-2}
\end{equation}
with Dirichlet boundary conditions on the lines (\ref{eq:3-2}).
Separating the coordinate $t$ from $\xi,\phi$ by the ansatz
\begin{equation}
\psi(t,\xi,\phi)=e^{-(i\omega+\fr)t}\psi(\xi,\phi)
\label{eq:4-3}
\end{equation}
and introducing the coordinates (\ref{eq:3-8}) we obtain
\begin{equation}
-y^2\left(\frac{\pat^2}{\pat x^2}+\frac{\pat^2}{\pat y^2}\right)
   \psi(x,y)=\left(\omega^2+\frac{1}{4}\right)\psi(x,y)
\label{eq:4-4}
\end{equation}
with Dirichlet boundary conditions on the lines (\ref{eq:3-9}). We
note that the eigenvalues $|\omega|$ give the allowed values of the
observable $|p_t|$ which classically equals the Lyapunov exponent and
the Kolmogorov Sinai entropy.

There is an immense mathematical literature on the eigenvalue problem
(\ref{eq:4-4}), see Hejhal~1976 \cite{m}, Helgason~1981 \cite{n},
Venkov~1982 \cite{o}. One important result is that the eigenvalues
$\omega$ of eq.  (\ref{eq:4-4}) are given exactly by the zeros
$s=\fr+i\omega$ of Selberg's zeta function, which is defined as an
infinite product over the primitive classical periodic orbits,
labelled by $p$, of geometric length $l_p$
\begin{equation}
  Z(s)=\prod_p\prod_k(1-\;\ep_k(p)\exp(-(s+k)l_p))
\label{eq:4-5}
\end{equation}
where $\ep_k(p)$ counts the number of reflections on the Dirichlet
boundaries (\ref{eq:3-9}) and is given by $\ep_k(p)=1$ if that number
is even, and $\ep_k(p)=-(-1)^k$ if that number is odd.

Another interesting observation, made by Bogomolny and Carioli~1992
\cite{b}, is the fact that the solution of eq. (\ref{eq:4-4}) can be
reduced to the eigenvalue problem of a 1-dimensional quantum map. In
general, the solutions $\psi(x,y)$ of eq. (\ref{eq:4-4}) have a
representation in terms of functions $f(t)$ on the real line
(Helgason~1981 \cite{n})
\begin{equation}
  \psi(x,y)=\int^{+\infty}_{-\infty}\,\left(\frac{y}{t^2+y^2}
     \right)^s f(x+t)dt.
\label{eq:4-6}
\end{equation}
which, like the $\psi$, transform according to an irreducible
representation of the Fuchsian group  $\Gamma$: $\psi_s(G(z))=
\chi(G)\psi_s(z)$; $f(G(t))|DG(t)|^{1-s}=\chi(G)f(t)$; $G\ep\Gamma$.
Demanding, e.g., that $\psi$ transforms according to the unit
representation of the Fuchsian group we find from (\ref{eq:3-10})
\begin{eqnarray}
   f(t) &=& f(t+2)\, ,\, f(t)=f(t-2)\nonumber\\
   f(t) &=& f\left(\frac{t}{2t+1}\right)|2t+1|^{-1+2i\omega}\\
   f(t) &=& f\left(\frac{t}{1-2t}\right)|1-2t|^{-1+2i\omega}.
            \nonumber
\label{eq:4-7}
\end{eqnarray}
The Dirichlet boundary conditions for $\psi$ are satisfied if $f(t)$
is restricted to odd functions
\begin{equation}
  f(t)=-f(t)
\label{eq:4-8}
\end{equation}
Thus, $\omega$ and $f(t)$ are determined as eigenpair from the
equation
\begin{equation}
  f(t) = f\left(\frac{t}{2t+1}(\mbox{\rm mod}\,2)\right)
    |2t+1|^{-1+2i\omega}
\label{eq:4-9}
\end{equation}
with the boundary condition (\ref{eq:4-8}). The 1-d map $t'=t/(2t+1)$
(mod 2) appearing in eq. (\ref{eq:4-9}) is similar to the Gauss map.
It appears as a one-dimensional map of the classical dynamics if the
transformations $A, B$ eq. (\ref{eq:3-10}) are restricted to the real
line. The `eigenfunctions' $f(t)$ are rather singular objects
(Helgason~1981 \cite{n}) which makes it preferable to study the
adjoint map induced by eq. (\ref{eq:4-9}) on the adjoints of $f(t)$
in a suitable scalar product. I refer to the papers by Bogomolny and
Carioli~1992 \cite{c}, Mayer~1991 \cite{p}, and Eisele and
Mayer~1993 \cite{f} for further discussion.

The methods described so far are interesting in their own right, but
they are not the most useful ones if one wishes to determine a large
number of eigenvalues with rather high precision. This is done best
by solving eq. (\ref{eq:4-4}) in a suitable representation directly
on a computer (Csord\'as et al~1991 \cite{a}, Schmit~1991 \cite{e}).
The general solution of eq. (\ref{eq:4-4}) satisfying Dirichlet
boundary conditions on $x=-1,0$ may be represented as
\begin{equation}
  \psi_\omega(x,y) =\sum^\infty_{n=1}A_n\sin(\pi nx)
   \sqrt{y}K_{i\omega}(\pi ny).
\label{eq:4-10}
\end{equation}
Here
\begin{equation}
  K_{i\omega}(z)=\fr\int_{-\infty}^\infty\, e^{-z\cosh t+i\omega t}
  dt.
\label{eq:4-11}
\end{equation}
The Dirichlet condition on the remaining side
$(x+\fr)^2+y^2=\frac{1}{4}$ yields
\begin{equation}
  0=\sum^\infty_{n=1}A_n\sin(\pi nx)
     \sqrt[4]{\frac{1}{4}-(x+\fr)^2}K_{i\omega}
       \left(\pi n\sqrt{\frac{1}{4}-(x+\fr)^2}\right).
\label{eq:4-12}
\end{equation}
The right-hand side can be expanded in the set $\{\sin(\pi
nx)\}^\infty_{n=1}$ which is complete in $[-1,0]$ and we obtain
\begin{equation}
  D_l=\sum_{n=1}^\infty \rmi_{ln} A_n =0
\label{eq:4-13}
\end{equation}
with
\begin{equation}
  \rmi_{ln}=
   4F_lG_n\int_{-1}^0\;dx\sin(\pi lx)\sin(\pi nx)
   \sqrt[4]{\frac{1}{4}-(x+\fr)^2}K_{i\omega}
   (\pi n\sqrt{\frac{1}{4}-(x+\fr)^2})
\label{eq:4-14}
\end{equation}
where $F_l$, $G_n$ can be chosen arbitrarily, not zero, to simplify
the eigenvalue condition
\begin{equation}
  \det \rmi_{ln}=0.
\label{eq:4-15}
\end{equation}
For the evaluation of the matrix elements (\ref{eq:4-14}) asymptotic
expansions of the Hankel functions are used, which are given in
Csord\'as et al~1991 \cite{a}. The eigenfunctions of the Hamiltonian
transform according to the irreducible representations of the
triangle group $C_{3v}$, which has two 1-dimensional irreducible
representations corresponding to non-degenerate states and one
two-dimensional representation. For the purposes of level statistics
each symmetry class must be considered separately. Eigenfunctions
transforming under the unit representation satisfy Neumann boundary
conditions on the symmetry axes, those transforming under the second
one-dimensional representation satisfy Dirichlet conditions there,
instead. For these cases the original triangle (\ref{eq:3-9}) is
replaced by the smaller triangle
\begin{equation}
  x=-\fr,\, x=0,\, y\ge 0;\, x^2+y^2=1,\, -\fr<x<0
\label{eq:4-16}
\end{equation}
i.e. in the Fourier expansions (\ref{eq:4-10})-(\ref{eq:4-13}) $n,l$
on the right hand side is replaced by $2n, 2l$ and
\begin{equation}
  \sqrt{\frac{1}{4}-(x+\fr)^2}\rightarrow\sqrt{1-x^2}.
\label{eq:4-17}
\end{equation}
In this way we restrict ourselves to a desymmetrized component of the
original energy spectrum.

Remarkably, the only reason for the non-integrability of the
Schr\"odinger eq. (\ref{eq:4-4}), (\ref{eq:4-16}) is the form of the
boundary at $y=\sqrt{1-x^2}$ which breaks the translational
invariance in $x$-direction. It is interesting therefore to unfold
the billiard problem and to consider not only the boundary
(\ref{eq:4-17}) but also the shifted boundary
\begin{equation}
  y=g(c_p,x)=1-\frac{1}{c_p}+
      \sqrt{\frac{1}{c_p^2}-x^2}
\label{eq:4-18}
\end{equation}
depending on the parameter $c_p$, $0\le c_p\le 1$. The boundaries
interpolate smoothly between the 'integrable' boundary $y=1$ for
$c_p=0$, which approximates (38) in the interval
$-\fr\le x\le 0$ reasonably well but preserves translational
invariance, and the real boundary (\ref{eq:4-16}) for $c_p=1$. The
integrable case $c_p=0$ has been investigated in Graham et al~1991
\cite{Z}, the general case $c_p<1$ and $c_p=1$ in Csord\'as et
al~1991 \cite{a} and, with emphasis on the transitions between the

different `universality classes', further in Csord\'as et al~1993
\cite{q}. We shall now proceed to discuss the level statistics
obtained in Csord\'as et al~1991 \cite{a}.

\section*{\bf LEVEL STATISTICS}

A very useful check on the completeness of the energy levels
$\lambda\equiv\omega^2+\frac{1}{4}$ obtained from eq. (\ref{eq:4-15})
up to a given energy $\lambda=\mu$ is provided by the smoothed level
density as obtained from the semi-classical Weyl formula and its
refinements. For the case at hand the Weyl formula was derived in
Csord\'as et al~1991 \cite{a}. If $\bar{N}(\mu)$ is the smoothed
number of eigenvalues $\lambda$ with $\lambda\le\mu$, the Weyl
formula reads
\begin{figure}[b]
\vskip14cm\caption[]{\small
The deviation of the spectral staircase from
the Weyl formula (a) $c_p=1$ and (b) $c_p=\fr$ (from Csord\'as et al
\cite{a}).}\end{figure}

\begin{equation}
 4\pi\bar{N}(\mu) = A\mu-\sqrt{\mu}\ln\mu-B\sqrt{\mu}+C+0(\mu^{-1/2})
\label{eq:5-1}
\end{equation}
with the area $A$ and the constants $B,C$ given as explicit functions
of $c_p$. The term $-\sqrt{\mu}\ln\mu$ is due to the diverging
circumference of the billiard. The numerically `measured' function
$N(\mu)$ deviates from $\bar{N}(\mu)$,
\begin{equation}
  N(\mu)=\bar{N}(\mu)+\fr-\delta N(\mu).
\label{eq:5-2}
\end{equation}
In figs.~3 a,b, we plot $\delta N(\mu)$ for $c_p=1$ and $c_p=\fr$
for about 1400 levels, respectively. It can be seen that $\delta
N(\mu)$ fluctuates around 0, indicating that no level has been missed
(a systematic shift of the average of $\delta N(\mu)$ by 1 would be
very noticeable in fig.~3). Another immediate observation from fig.~3
is that the fluctuations are significantly {\bf larger} in the case
$c_p=1$ than in the case $c_p=1/2$. Using the smoothed
level-staircase (\ref{eq:5-1}) the average level density can be
scaled out by the transformation
\begin{equation}
  x_i=\bar{N}(\lambda_i).
\label{eq:5-3}
\end{equation}
Then it is possible to determine the nearest neighbor level-spacing
distribution $P(s)ds$ which gives the probability to find the next
level $x_{i+1}$ in the interval $(x_i+s,x_i+s+ds)$ if there is a
level at $x_i$. The integrated level spacing distributions
$\rmi(s)=\int_0^s\;P(s)ds$ for $c_p=1,\fr$ are plotted in fig.~4 a,b,
respectively, and compared with the Poissonian result (dashed curve)
\begin{equation}
 \rmi_P(s)=1-e^{-s}
\label{eq:5-4}
\end{equation}
and the Wigner surmise for the Gaussian Orthogonal Ensemble (GOE) of

random matrix theory (dashed-dotted-curve)
\begin{equation}
 \rmi_{W}(s)=1-e^{-\pi s^2/4}.
\label{eq:5-5}
\end{equation}
For $c_p=1/2$ there is good agreement with the GOE result, indicating
that the transition from the integrable to the chaotic case has
occurred. However, the case $c_p=1$ fits the Poissonian distribution
better than the GOE distribution, which was  surprising at first (see
Balazs and Voros~1986 \cite{Q} for the first observation of this
phenomenon in a compact billiard), indicating some hidden non-generic
behavior in this prototypical chaotic system. This non-genericity is
now understood to be a consequence of hidden
\begin{figure}[b]\vskip14cm\caption[]{\small
The integrated nearest-neighbor level spacing distribution
from approximately 1400 levels for (a) $c_p=1$ and (b) $c_p=\fr$.
Solid, dashed, and dotted lines are the numerical result, the
Poissonian, and the GOE predictions, respectively (from
Csord\'as et al \cite{a}).}
\end{figure}
degeneracies occurring in some billiards on the pseudo-sphere which
are due to certain special arithmetical properties of the underlying
Fuchsian group (see below).

In fig.~5 a,b we also plot the averaged spectral rigidity
\begin{equation}
  \bar{\Delta}_3(L,x)=\frac{1}{p}\int_{x-p/2}^{x+p/2}\,
     \Delta_3(L,x')dx'
\label{eq:5-6}
\end{equation}
with
\begin{equation}
  \Delta_3 (L,x)=\min_{A,B}\frac{1}{L}\int_{x-L/2}^{x+L/2}\,
   (N(x')-Ax'-B)^2 dx'
\label{eq:5-7}
\end{equation}
again for $c_p=1, 1/2$. For comparison the Poisson statistics
$\bar{\Delta}_3=L/15$ and the GOE statistics
$\bar{\Delta}_3\simeq\frac{1}{\pi^2}\ln L-0.00695\dots$ for large $L$
are also shown. For $c_p=1/2$ the GOE result fits the data well,
while for $c_p=1$ the $\bar{\Delta}_3$-statistics for small $L$ is
again closer to the Poissonian result. For larger $L$ the Poissonian
result is not expected to be a good approximation, not even in the
integrable case, because of the influence of periodic orbits
(Berry~1985 \cite{r}). For the integrable case $c_p=0$ see Graham et
al~1991 \cite{Z}. The conclusion that the case $c_p=1$ is non-generic
and similar to an integrable case is further confirmed by the
$\Sigma_2$-statistics
\begin{equation}
  \Sigma_2=\langle [n_i(r)-\langle n_i(r)\rangle]^2\rangle
\label{eq:5-8}
\end{equation}
where $n_i$ is the number of levels in an interval of size $r$. These
data are shown in figs.~5 a, b respectively. An explanation for the
strong deviation from GOE statistics in the chaotic case $c_p=1$ has
been given by Bolte et al~1992 \cite{d} and Bogomolny et al~1992
\cite{b}. It is due to the existence of further symmetries of some of
the Fuchsian groups (called `arithmetic'), to which the infinite
equilateral triangle also belongs. Classically these additional
symmetries give rise to the appearance of a multiplicity $\langle
g\rangle$ of periodic orbits of the same length $l$ with $\langle
g\rangle$ for length $l<M$ growing exponentially
with $M/2$ for large $M$
\begin{equation}
 \langle g\rangle=\frac{\sum_{l<M}g(l)}{\sum_{l<M}}= \const
   \frac{e^{M/2}}{M}.
\label{eq:5-9}
\end{equation}
When used, in Berry's~1985 \cite{r} semi-classical theory of level
statistics the large value of $\langle g\rangle$ for $M>>1$ has
consequences very similar to those of the infinite multiplicities of
periodic orbits in integrable systems. Quantum mechanically for
arithmetic Fuchsian groups there exists an infinite denumerable
set of hermitian operators $T_n$ `Hecke operators', commuting among
themselves and with $H$, which can be diagonalized simultaneously
with $H$ (Hejhal~1983 \cite{m}, Venkov~1990 \cite{s}).  Again one
then expects Poissonian statistics, as observed. Even a slight shift
of one of the walls $(c_p<1)$ is enough to destroy the
length-degeneracy of the periodic orbits (or the conservation of the
Hecke operators) and leads to a transition to GOE statistics for
sufficiently high-lying levels. This transition has recently been
studied in detail by Csord\'as et al~1993 \cite{q}, who found a
scaling behavior in the variables $1-c_p$ and the energy quantum
number $n$.

\begin{figure}[t]\vskip14cm\caption[]{\small
The spectral rigidity $\Delta_3(L,x)$ at $x=700$
for (a) $c_p=1$ and (b) $c_p=\fr$. Solid, dashed, and dotted lines
are the numerical result, the Poissonian, and the GOE predictions,
respectively (from Csord\'as et al \cite{a}).}\end{figure}

\begin{figure}[t]\vskip14cm\caption[]{\small
$\Sigma_2(r)$ statistics for (a) $c_p=1$ and (b)
$c_p=\fr$.  Solid, dashed, and dotted lines are the numerical result,
the Poissonian, and the GOE predictions, respectively (from Csord\'as
et al \cite{a}).}\end{figure}

\section*{\bf SEMI-CLASSICAL SOLUTIONS OF THE WHEELER DEWITT EQUATION}

The study of the separated eigenvalue euqation (\ref{eq:4-4})
presented in the two preceding sections has several drawbacks: (1) It
can claim validity only in the asymptotic region
$t\rightarrow+\infty$ where $t$ is the function of
$\alpha,\gb_+^2+\gb_-^2$ defined in eq. (13). (2) It
completely neglects the influence of the potential channels at the
corners of the triangle in configuration space, which become
extremely narrow for $t\rightarrow+\infty$, but, in principle remain
always open. (3) Even if eq. (\ref{eq:4-4}) gave a complete
description it would leave open the question which linear
superposition of the eigenstates of (\ref{eq:4-4}) determines the
actual wave-function of the Universe.

For all these reasons it seems desirable to avoid the transformation
(13) and instead to go back to the original Wheeler DeWitt
equation (\ref{eq:4-1}), even at the cost of treating a more
difficult problem.

As we are still interested in the asymptotic region where the
influence of matter is negligible we may anticipate that the
solution of (\ref{eq:4-1}) will have the form of a bound state
\begin{equation}
  \psi(\alpha,\gb_+,\gb_-)\sim \exp
(-\phi(\alpha,\gb_+,\gb_-)/\hbar)
\label{eq:6-1}
\end{equation}
where we restore Planck's constant here and in (\ref{eq:4-1}) in
order to have a formal control over the semiclassical limit
$\hbar\rightarrow0$. Surprisingly, exact analytical solutions of the
Wheeler DeWitt equation of the form (\ref{eq:6-1}) can be given
(Graham 1991 \cite{h}, Moncrief and Ryan~1991 \cite{g} Bene and
Graham~1993 \cite{i}). The potential $\phi$ in eq.~(\ref{eq:6-1}) may
be interpreted as a Euclidean action which is required for the system
in its bound state to tunnel from some initial point
$\alpha_0,\gb_{+0}, \gb_{-0}$ to a given end-point
$\alpha,\gb_+,\gb_-$. (In this context it may be helpful to recall
a harmonic oscillator with coordinate $x$ in its ground state where
in appropriate units $\phi=x^2$ is the Euclidean action for
tunnelling from $x=0$ to the classically forbidden amplitude
$x\neq 0$).

Inserting the ansatz (\ref{eq:6-1}) in eq. (\ref{eq:4-1}) (and taking
$\hbar\rightarrow 0$ or adding suitable terms of order $\hbar$ to the
potential $V$ which is always possible by an appropriate choice of
factor ordering when defining the quantized model, See Moncrief and
Ryan~1990 \cite{f}) we find the
Euclidean Hamilton Jacobi equation \footnote{A Euclidean Hamilton
Jacobi equation (\ref{eq:6-2}) with a potential
$V_0(\alpha,\gb_+,\gb_-)$ via eq. (\ref{eq:6-1}) corresponds to a
Wheeler DeWitt equation (\ref{eq:4-1}) with a potential
\[
  V=V_0(\alpha,\gb_+,\gb_-)+\frac{\hbar}{2}
\left( \frac{\pat^2\phi}{\pat\alpha^2}-
        \frac{\pat^2\phi}{\pat\gb_-^2}-
\frac{\pat^2\phi}{\pat\gb^2_-}\right)
\]
Such a potential appears naturally in supersymmetric formulations
(Graham 1991 \cite{g}, Bene and Graham~1993 \cite{j}). However, in
any case $V$ is reduced to $V_0$ in the semiclassical limit
$\hbar\rightarrow 0$.}
\begin{equation}
  -\fr\left(\frac{\pat\phi}{\pat\alpha}\right)^2+
\fr\left(\frac{\pat\phi}{\pat\gb_+}\right)^2+
\fr\left(\frac{\pat\phi}{\pat\gb_-}\right)^2- V(\alpha,\gb_+,\gb_-)=0
\label{eq:6-2}
\end{equation}
which now contains the {\bf inverted} potential $-V$. The following
exact analytical solutions of this Euclidean Hamilton Jacobi equation
are known, Belinsky et al~1978 \cite{t}, Gibbons and Pope~1979
\cite{u}, Atiyah and Hitchin~1985 \cite{v}, in the case of Bianchi
type IX
\begin{eqnarray}
 \phi_1 &=& \frac{1}{6} \left(
e^{2\gb_++2\sqrt{3}\gb_-}+e^{2\gb_+-\sqrt{3}\gb_-}+e^{-4\gb_+}
\right)\nonumber\\ \phi_2 &=& \phi_1-\frac{1}{3}e^{2\alpha} \left(
e^{2\gb_+}+e^{-\gb_++\sqrt{3}\gb_-}+e^{-\gb_+-\sqrt{3}\gb_-}
\right)\nonumber\\ &=&
V\left(\frac{\alpha}{2},\frac{\gb_+}{2},\frac{\gb_-}{2}
\right)\nonumber\\ \phi_3 &=& \phi_1-\frac{1}{3}e^{2\alpha} \left(
e^{2\gb_+}-e^{-\gb_++\sqrt{3}\gb_-}-e^{-\gb_+-\sqrt{3}\gb_-} \right).
\label{eq:6-3}
\end{eqnarray}
{}From $\phi_3$ two other solutions may be obtained by $120^\circ$
degree rotations in the $(\gb_+,\gb_-)$-plane. It is clear that
$-\phi$ is also a solution if $\phi$ is a solution. In the following
we shall only consider the solutions $\phi_1,\phi_2$ which have the
full triangular symmetry of the problem. In figs.~6 and 7 we give
plots of the wave-functions (\ref{eq:6-1}) obtained with
$\phi_1,\phi_2$ for different fixed values of $\alpha$. The maximum
$\psi$ is always normalized to unity. It is seen that the channels
indeed modify the wave-function. However, because the $\phi_i$ are
real there is only a tunnelling penetration into the channels
without any real propagation inside them.

\dib{22cm}{Wormhole state: The wave function $|\psi|^2$ for
$\phi=\phi_1$ as a function of $\gb_+,\gb_-$ for four
different values of $\alpha$. The maximum of $|\phi|^2$ is always
normalized to 1. The changing scale of $\gb_+,\gb_-$ should be noted
(after Graham~1991 \cite{h}).}

\dib{22cm}{No-boundary (Hartle- Hawking) state: The same as
Fig.~7 but for $\phi=\phi_2$.}

For the case of Bianchi type VIII very similar
analytical solutions exist. They are given by (Bene and Graham~1993
\cite{i})
\begin{eqnarray}
  \phi_1 &=& \frac{1}{6}e^{2\alpha}
   \left(
    e^{2\gb_++2\sqrt{3}\gb_-}+e^{2\gb_+-2\sqrt{3}\gb_-}-e^{-4\gb_+}
   \right)\nonumber\\
  \phi_2 &=& \frac{1}{6}e^{2\alpha}
   \left(
    4e^{2\gb_+}\sinh^2\sqrt{3}\gb_--e^{-4\gb_+}\pm4ie^{-\gb_+}
        \cosh\sqrt{3}\gb_-
    \right)\\
     \phi_3 &=& \frac{1}{6}e^{2\alpha}
   \left(
     4e^{2\gb_+}\cosh^2\sqrt{3}\gb_--e^{-4\gb_+}\pm4ie^{-\gb_+}
        \sinh\sqrt{3}\gb_-
   \right)\nonumber
\label{eq:6-4}
\end{eqnarray}
The fact that $\phi_2$ and $\phi_3$ turn out to be complex is due to
the non-binding nature of the Bianchi VIII potential which now makes
a real propagation of the wave-function (without tunnelling) inside
the channel along the positive $\gb_+$-axis possible and gives rise
to a real physical action (i.e. an imaginary Euclidean action).  For
$\alpha\rightarrow-\infty$ and values of $\gb_+,\gb_-$ inside the
triangle (\ref{eq:2-11}) the imaginary part of $\phi$ is negligibly
small. Yet solutions of the form (\ref{eq:6-1}) with
$\phi=\pm\phi_j$, $\phi_j$ of eq. (53), turn all out to be
non-normalizable for fixed $\alpha$, i.e. none of these solutions is
physically acceptable for Bianchi type VIII.

We now turn to the problem of physically interpreting the special
solutions (\ref{eq:6-3}). First of all, this involves the question of
the interpretation of the wave-function of the Universe, which would
need a discussion we don't wish to enter here (see e.g. the
discussion in Bene and Graham~1993 \cite{i}). We shall simply
interpret $|\psi|^2$ for fixed logarithm of the scale parameter
$\alpha$, as the probability distribution of the anisotropy
parameters $\gb_+,\gb_-$ for that scale parameter. However, we still
have to understand which wave-function applies to which physical
situation. This question can be answered by recalling the meaning of
$\phi$ as a Euclidean action. The (imaginary-time) solutions of the
classical equations of motion corresponding to these actions can be
interpreted (for $\hbar\rightarrow0$) as the most probable paths
followed in the tunnelling propcess. These paths, in the present
case, correspond to a sequence of 3-spaces followed in the tunnelling
process from a given intial 3-space up to the final 3-space whose
parameters $\alpha,\gb_+,\gb_-$ appear as the arguments of $\phi$.
The initial point of the tunnelling path is chosen as a most probable
state $\phi=$min for $\alpha$ fixed where the system likes to sit
when it does not fluctuate away on one of its tunnelling excursions
to the neighborhood (again, recall the harmonic oscillator in its
ground state for a simple example of this physical picture). Let us
now consider the most likely tunnelling paths. They satisfy
\begin{eqnarray}
  p_\alpha &=& -\frac{d\alpha}{d\lambda}=
          \frac{\pat\phi}{\pat\alpha}=2\phi\nonumber\\
p_+       &=&
          \frac{d\gb_+}{d\lambda}= \frac{\pat\phi}{\pat\gb_+}\\
p_-       &=&
          \frac{d\gb_-}{d\lambda}= \frac{\pat\phi}{\pat\gb_-}
          \nonumber
\label{eq:6-5}
\end{eqnarray}
It follows that
\begin{eqnarray}
  \frac{d\gb_+}{d\alpha} &=& -\fr\frac{\pat}{\pat\gb_+}
U(\gb_+,\gb_-)\nonumber\\ \frac{d\gb_-}{d\alpha} &=&
-\fr\frac{\pat}{\pat\gb_-} U(\gb_+,\gb_-)
\label{eq:6-6}
\end{eqnarray}
with the new $\alpha$-independent potential
\begin{equation}
  U(\gb_+,\gb_-) = \ln|e^{-2\alpha}\phi(\alpha,\gb_+.\gb_-)|.
\label{eq:6-7}
\end{equation}
According to eq. (\ref{eq:6-6}) the evolution of $\gb_+,\gb_-$ with
increasing $\alpha$ is on lines of steepest descent of the potential
$U$.

Let us now consider the concrete  Bianchi type IX solutions $\phi_1$
and $\phi_2$ of eq. (\ref{eq:6-3}). For $\phi=\phi_1$ eqs.
(\ref{eq:6-5}) can be solved explicitely (Belinsky et al~1978
\cite{t}) yielding the Riemannian space-time metric
\begin{equation}
  ds^2=
   \frac{d\rho^2}{\sqrt{F(\rho)}}+\frac{1}{4}\rho^6\sqrt{F(\rho)}
   \left(
    \frac{\omega^1\omega^1}{\rho^4-a_1^4}+
    \frac{\omega^2\omega^2}{\rho^4-a_2^4}+
    \frac{\omega^3\omega^3}{\rho^4-a_3^4}
   \right)
\label{eq:6-8}
\end{equation}
with
\begin{equation}
  F(\rho)=\left(1-\frac{a_1^4}{\rho^4}\right)
  \left(1-\frac{a_2^4}{\rho^4}\right)
  \left(1-\frac{a_3^4}{\rho^4}\right)
\label{eq:6-9}
\end{equation}
where $a_1,a_2,a_3$ are constants of integration which must be chosen
in such a way that the spatial 3-metric of eq. (\ref{eq:6-8}) becomes
equal to the Bianchi type IX 3-metric with given $\alpha,\gb_+,\gb_-$
for a suitable choice of $\rho$. It follows from (\ref{eq:6-8}) that
the Euclidean coordinate time $(it)$ is
$d(it)=(F(\rho))^{-1/4}d\rho$. The solution (\ref{eq:6-8}) must be
restricted to $\rho^2>\max(a_1^2,a_2^2,a_3^2)$ otherwise the
3-geometry is separated from the initial 3-geometry by a singularity.

The tunnelling path starts at the
minimum of $\phi_1$ for fixed $\alpha$, which is at $\gb_+=0=\gb_-$,
corresponding to $\rho=\infty$ in eq. (\ref{eq:6-8}). The given
Bianchi-IX metric corresponds to the {\bf inner} boundary at small
$\rho^2$ (but $\rho^2>\max (a_1^2, a_2^2, a_3^2)$). For
$\rho\rightarrow\infty$ the 4-metric (\ref{eq:6-8}) approaches
$ds^2=d\rho^2+\frac{1}{4}\rho^2(\omega^1\omega^1+
\omega^2\omega^2+\omega^3\omega^3)$,i.e. it becomes Euclidean and
flat (Belinsky et al~1978 \cite{t}), indicating that the
wave-function under discussion here describes the spontaneous
quantum fluctuation of a flat 3-metric to one with an anisoptropic
inner boundary or throat, i.e. a quantum wormhole
(D'Eath~1993 \cite{j}, Bene and Graham~1993 \cite{i}).

Turning now to $\phi=\phi_2$ we note that a transparent closed-form
solution like (\ref{eq:6-8}) is not available in this case. However,
we may use eqs. (\ref{eq:6-6}) and the first of eqs. (54)
to obtain the complete physical picture. Let us note that for
$\phi=\phi_2$ the equi-potential lines of $\phi_2$ and of
$U(\gb_+,\gb_-)$ coincide with those of $V(\alpha,\gb_+,\gb_-)$ for
fixed $\alpha$. The minimum of $\phi_2$ for fixed $\alpha$ is again
at $\gb_+=0=\gb_-$ defining the most likely starting point of the
tunnelling path. This point corresponds to a local isotropic maximum
of $U$, i.e. if $\gb_+,\gb_-$ beginn to tunnel to small non-zero
values, $\alpha$ increases from its inital value $-\infty$ meaning
that the Universe expands from nothing in some random anisotropic
way. From the first of eqs. (54) it follows that this
expansion occurs with increasing $\lambda$, since
$\phi_2(\alpha,0,0)<0$. However, $U(\gb_+,\gb_-)$ has a degenerate
minimum on the line where $V(\alpha,\gb_+,\gb_-)$ (and hence also
$\phi_2$) change sign. In fact this line is formed by a triangle of
three hyperbolas which are asymptotic to the positive $\gb_+$ axis
and the two axes obtained by $\pm 120^\circ$ rotations. Outside this
triangle of hyperbolas the equipotential lines of $V$, $\phi_2$ and
$U$ are no longer closed due to the existence of the channels on the
three axes. Linearizing eqs. (\ref{eq:6-5}) in the neighborhood of
the line of minimal $U$ one finds that with increasing $\lambda$ the
line is in fact crossed, but while crossing $\alpha$ reaches a
maximum and decreases again as $\gb_+,\gb_-$ start climbing the
potential $U(\gb_+,\gb_-)$ outside the triangle of hyperbolas to
their given final value.

Thus for given values of $(\gb_+,\gb_-)$ outside (or inside) the said
triangle of hyperbolas the most probable path of tunnelling starts
with vanishing 3-volume ($\alpha=-\infty$) and goes (or,
respectively,  does not go) through a maximum of the 3-volume until
it reaches the given values $\alpha,\gb_+,\gb_-$. The given 3-metric
in this case forms the {\bf outer} (and in fact the only) boundary of
the compact Riemannian space-time region traced out by the
3-geometries along the most probable tunnelling path, and $\phi_2$ is
the Euclidean action of this compact space-time region. Hartle and
Hawking~1983 \cite{S} made the general proposal that in the
semi-classical limit the wave-function of the universe is given as
the exponential of the Euclidean action constructed in this way. Thus
\begin{equation}
  \psi=\const\exp(-\phi_2(\alpha,\gb_+,\gb_-)/\hbar)
\label{eq:6-10}
\end{equation}
is identified as the no-boundary state of the Bianchi-type IX quantum
space-time (Bene and Graham~1993 \cite{i}). The state has also been
discussed by Hawking and Luttrell~1984 \cite{T} and by Moss and
Wright~1985 \cite{U}, however, this work was qualitative or
numerical and the analytical expression (\ref{eq:6-10}) was not given
there.

A similar discussion for $\phi_3$ of eq.~(\ref{eq:6-3}) shows that the
wave-function (\ref{eq:6-1}) for this case arises by quantum
tunnelling from a disk shaped 3-geometry ($\gb_+=\infty$, $\gb_-=0$)
with infinite volume ($\alpha\rightarrow\infty$) down to a given
3-geometry $\alpha,\gb_+,\gb_-$ with $\alpha$ decreasing monotonously
along the most probable tunnelling path which closely follows the
$\gb_+$-axis before it finally turns to the given value $\gb_-$. This
is again a solution of wormhole type, but this time of a virtual
wormhole in a highly anisotropic 3-geometry.

Finally, let us compare the solutions obtained in the present section
with the solutions obtained in the two preceding sections. First, we
note that the solutions given there have only restricted
validity, as they were based on an approximation which is only  valid
in the limit $\alpha\rightarrow-\infty$. On the other hand, in that
limit, the Wheeler DeWitt equation, has been solved there, in
principle, as a {\bf general} linear superposition of the solutions
of the eigenvalue problem of the two-dimensional billiard. In order
to select the special cosmologically relevant solution an additional
`initial' or boundary condition would still be needed. The simplest
choice is to select the ground state of the billiard (Graham and
Sz\'epfalusy~1990 \cite{W}), but other choices may, of course, be
possible. By contrast, in the present section we have
obtained some particular solutions only, which are, however, valid
for arbitrary $\alpha$. As discussed above, these solutions
correspond to the special boundary conditions of the `no-boundary
state' and the wormhole state. Their cosmological significance is
therefore clear. If specialized to the regime
$\alpha\rightarrow-\infty$ these particular solutions may be
represented as linear superpositions of the eigenmodes of the
two-dimensional billiard, i.e. they can be used to determine the
probability amplitudes of the eigenmodes of the billiard in the
wave-function of the Universe. As can be seen from figs.~6, 7 for
$\alpha\rightarrow-\infty$ not only the ground state of the billiard,
but also arbitrarily high lying states will be excited, due to the
increasingly sharp drop of the wave-function from a nearly constant
value inside the billiard to 0 at the boundary. In fact it appears
that these solutions asymptotically approach the constant solution of
the billiard for Neumann boundary conditions which contains
arbitrarily high lying states satisfying Dirichlet boundary
conditions. The kinetic energy of
the billiard is the positive energy residing in the anisotropy
degrees of freedon $\gb_+, \gb_-$ or $x,y$. This positive energy is
exactly balanced, because of $H=0$, by the negative energy
$-\fr p_t^2$ of gravitational attraction. The particular solutions
discussed in the present section, in the limit
$\alpha\rightarrow-\infty$, contain components for which both
energies are unbounded.

\section*{\bf ACKNOWLEDGEMENT}

I am very greatful to Peter Sz\'epfalusy for first stimulating my
interest in the quantum mechanics of the mixmaster cosmology.  The
results on the level statistics were obtained in joint work with
P.~Sz\'epfalusy, A.~Csord\'as, G.~Vattay and R.~H\"ubner.  The work
reported in the last section was done, in part, in collaboration with
J.~Bene. All those mentioned I would like to thank for a most
enjoyable and fruitful collaboration. Finally I would like to thank
Peter D'Eath and Gary Gibbons for enlightening discussions concerning
the topics of the last section.

This work was supported by the Deutsche Forschungsgemeinschaft
through the Sonderforschungs\-bereich 237 `Unordnung und gro{\ss}e
Fluktuationen' and by the `Deutsch-Ungarisches Kooperationsabkommen'
through grant X231.3.

\end{document}